\documentclass[conference]{IEEEtran}
\usepackage{graphicx} 
\graphicspath{{plots/}}
\usepackage{listings}
\usepackage{xcolor}   
\usepackage{flafter}

\usepackage{algorithm} 
\usepackage{algpseudocode} 
\usepackage{multirow}
\usepackage{amsmath}
\usepackage{caption}
\usepackage{subcaption}
\usepackage{authblk}
\usepackage[nolist]{acronym}

\begin{acronym}
    \acro{IoT} {Internet of Things}
    \acro{YAFS} {Yet Another Fog Simulator}
    \acro{RR} {Round Robin}
    \acro{BW} {Bandwidth}
    \acro{PD} {Propagation Delay}
    \acro{GW} {Gateway}
    \acro{IPT} {Instructions per Simulation Time}
    \acro{QoS} {Quality of Service}
\end{acronym}

\begin{document}

\title{Optimizing Microservices Placement in the Cloud-to-Edge Continuum: A Comparative Analysis of App and Service Based Approaches}

\author[1]{Miguel Mota-Cruz}
\author[1]{João H. Santos}
\author[1]{José F. Macedo}
\author[1]{Karima Velasquez}
\author[2,1]{David Perez Abreu}

\affil[1]{Centre for Informatics and Systems of the University of Coimbra}
\affil[2]{Instituto Pedro Nunes}

\maketitle

\begin{abstract}
In the ever-evolving landscape of computing, the advent of edge and fog computing has revolutionized data processing by bringing it closer to end-users. While cloud computing offers numerous advantages, including mobility, flexibility and scalability, it introduces challenges such as latency. Fog and edge computing emerge as complementary solutions, bridging the gap and enhancing services' proximity to users. The pivotal challenge addressed in this paper revolves around optimizing the placement of application microservices to minimize latency in the cloud-to-edge continuum, where a proper node selection may influence the app's performance. Therefore, this task gains complexity due to the paradigm shift from monolithic to microservices-based architectures. Two distinct placement approaches, app-based and service-based, are compared through four different placement algorithms based on criteria such as link latency, node resources, and gateway proximity. App-based allocates all the services of one app sequentially, while service-based allocates one service of each app at a time. The study, conducted using YAFS (Yet Another Fog Simulator), evaluates the impact of these approaches on latency and load balance. The findings consistently confirm the hypothesis that strategies utilizing a service-based approach outperformed or performed equally well compared to app-based approaches, offering valuable insights into trade-offs and performance differences among the algorithms and each approach in the context of efficient microservices placement in cloud-to-edge environments.
\end{abstract}

\section{Introduction}
\label{sec:intro}
A transition from centralized cloud architectures to a distributed model encompassing cloud, fog and edge computing has led to a profound paradigm shift in the landscape of computing. This shift has been driven by the escalating demand for real-time processing, lower latency, and enhanced system performance \cite{article}. 
Although this integration offers extensive connectivity, it originates new challenges in data processing, management and security. In order to mitigate those issues, Fog and Edge Computing have emerged as strategic responses, which extend the capabilities of Cloud Computing to the lower layers of a network, thereby facilitating real-time data processing, and reducing latency \cite{DAS2023100049, article4}. 
Alongside this paradigm shift, the cloud-to-edge continuum \cite{9237349} has become a key concept, representing a spectrum of computing resources and network latency that vary across the different layers. In the upper layer, there is cloud computing, characterized by high computing resources and high latency. On the opposite side of the spectrum, the collection of \acp{GW} forms the edge layer, characterized by low latency and processing resources. The fog layer is constituted by the network nodes in between the cloud and \ac{GW} nodes. 
As one descends through the layers, from the cloud towards the edge, the computing resources of the nodes frequently decrease, but with the counter benefit of a reduction in latency. This reduction occurs from the increasing proximity to the \acp{GW}, which serves as an interface between the end-users and the continuum.
While the differences in resources and latency between two consecutive layers in this continuum might be subtle, the contrast becomes stark when comparing the extremes: a cloud and an edge node.
In parallel with these developments, there has been a shift in application architecture and development from monolithic to microservices-based applications. In contrast to monolithic architectures, where applications are single indivisible units, microservices architecture breaks applications down into smaller, functionally independent services, which offers enhanced scalability and flexibility as well as easier updates and maintenance, in benefit of cloud-to-edge environments \cite{9235316}. 
Furthermore, it provides increased resilience by allowing partial failures instead of the entire application failing.
Within this evolving paradigm, fog and edge computing and microservices, the strategic placement of applications services across the cloud-to-edge continuum emerges as a critical factor in optimizing system performance. As addressed in this paper, the dare lies in the determination of the most effective placement of application microservices to optimize latency, a crucial aspect in real-time data processing in the continuum. The investigation that has been carried on delves into this challenge, exploring different placement algorithms and their impact on latency and network load balance. Bearing this goal in perspective, two primary approaches have been taken into focus: app-based and service-based placements. In app-based placement, all the services of one app are placed, preferably, on the same node, before proceeding with the following application. On the other hand, service-based placement performs the allocation of one service of each application at a time as a \ac{RR} through the services of each application.
This research aims to evaluate the performance of these two approaches continuum environment context. By conducting simulations and analyses, the intention is to provide insights into which approach, app-based or service-based, effectively minimizes latency. The thesis hypothesizes that, in the context of high-load scenarios, service-based placement will demonstrate a superior capacity in reducing overall latency, thereby ensuring a more consistent application performance across the scenario and an overall lower response time for the end-users. The rationale behind this hypothesis lies in the limitation of node resources inherent in fog and edge computing networks. App-based placement, while optimizing latency for individual applications by attempting to allocate all its services in the same node, as node capacity exhausts, significant challenges arise, leading to the subsequent applications being pushed to nodes farther away from the edge, increasing the latency of that application. On the contrary, service-based placement, which sequentially allocates one service of each application before moving to the next, provides a more distributed spread of services over the topology. As the network load intensifies, this method is postulated to maintain a more uniform performance across all applications, avoiding the scenario where a few applications benefit from optimized performance at the expense of lower performance for others.
The subsequent sections of this article delve into related work, exploring fog and edge computing, microservices, and their strategic placements. The methodology outlines simulation steps, parameters and algorithms. Following, the findings are presented in the results section, offering a comparative analysis of app-based and service-based placements in the cloud-to-edge continuum, interpreted in the discussion for system performance optimization. Finally, the conclusion encapsulates the key findings, affirming the hypothesis and suggesting future exploration.



\section{Related Work}
\label{sec:rel_work}

Fog computing addresses challenges related to the performance of microservices applications, offering solutions to various issues. Extensive work has been conducted in this domain to enhance system efficiency and overcome associated problems, including microservice placement. Regarding this, the authors in \cite{8255573} have proposed a model for scheduling microservices over heterogeneous cloud-edge environments.
Alternative approaches have been explored in \cite{article2}, where authors have presented a heuristic solution based on a genetic algorithm that maximizes the usage of fog devices, resulting in a reduction of network communication delays. Moreover, this matter is addressed in \cite{9235316} through a profiling-based microservices placement algorithm that resorts to profiling experiments with selected workloads and a greedy-based heuristic algorithm, utilizing resource requirements derived from the profiled results.
Contributing with a heuristic algorithm, \cite{AZIZI20249} proposes the delay and cost-aware service placement that takes into consideration delay and cost of services, aiming to improve utilization of resources, delay, and cost for delay-sensitive services. 
In comparison to the existing body of work mentioned in the related section, this research builds upon these efforts by focusing on the specific context of cloud-to-edge continuum environments. While previous studies have made significant contributions to microservices placement in fog and edge computing, our work further extends this exploration by evaluating the impact of two different placement approaches (app-based and service-based) through diverse strategic algorithms. By conducting simulations and analyses using \ac{YAFS}, the aim is to provide insights into the trade-offs and performance differences among these approaches and strategies. Additionally, this study encompasses a comprehensive evaluation of both latency and load balance, offering a holistic understanding of the placement approaches effectiveness.
This contributes to the ongoing discourse on efficient microservices placement, especially in high-load scenarios involving real-time data processing and the evolving landscape of fog and edge computing.














\section{Methodologies}
\label{sec:meth}



The exploration of application performance within the cloud-to-edge continuum has been carried out with a simulator tool. The resort to a simulator, in spite of a testbed, is justified by the ease of implementation and test scalability.
In this context, \ac{YAFS} emerges as a versatile and customizable solution.
\ac{YAFS} Incorporates several functionalities and customizations based on its classes \cite{8758823}. \ac{YAFS} also provides a range of metrics, including latency, network usage and service response time. 
Its ability to model a variety of scenarios and assess the performance of different service placement strategies aligns closely with the objectives of this research, which focuses on exploring the comparative effectiveness of app-based versus service-based placements in minimizing latency in \ac{IoT} environments.
The parameters provided in Table \ref{tab:scenario_configuration} were used, which are similar to the values applied in previous work \cite{8588297}. The network setup involves 100 nodes, each with varied resources and processing speed. The cloud is modelled with substantially higher capacity and processing power. For the applications, 20 different apps were simulated with varying numbers of microservices. This configuration aims to provide a comprehensive understanding of the impact of the placement approaches on latency in a fog and edge computing environment, ensuring that the findings are applicable to real-world scenarios. 
A key component of this analysis, and therefore of this simulation, is the placement strategy, for which four distinct placement algorithms have been developed and incorporated in the simulation. Furthermore, each algorithm simulation has been run 50 times to provide more robust and reliable results.

\begin{table}[t]
\captionof{table}{Scenario Parameters Configuration}
\centering
\begin{tabular}{| c | c | c |}
 \hline
 Element & Parameters & Values\\ 
 \hline \hline
 \multirow{5}{*}{Network}  & NUMBER\_OF\_NODES & 100 \\ 
                           & NODE\_RESOURCES & 10-25 MB RAM \\
                           & NODE\_SPEED & 500-1000 Instructions/ms \\
                           & LINK\_BANDWIDTH & 75000 Bytes/ms \\
                           & PROPAGATION\_DELAY & 2-10 ms \\
\hline
 \multirow{4}{*}{Cloud}    & CLOUD\_CAPACITY & $\infty$ \\ 
                           & CLOUD\_SPEED & 10000 Instructions/ms \\
                           & CLOUD\_BANDWIDTH & 125000 Bytes/ms \\
                           & PROPAGATION\_DELAY & 500 ms \\
\hline
 \multirow{2}{*}{APPs}     & NUMBER\_OF\_APPS & 20 \\ 
                           & NUMBER\_OF\_SERVICES\_p\_APP & 2-10 \\
\hline
\multirow{3}{*}{Modules}   & MODULE\_INSTRUCTIONS & 20000-60000 \\ 
                            & MODULE\_RESOURCES & 1-6 MB RAM \\
                            & MODULE\_MESSAGE\_SIZE & 1500000-4500000 Bytes \\
\hline
User & REQUEST\_RATE & 200-1000 \\ 
\hline

\end{tabular}
\label{tab:scenario_configuration}
\end{table}


\section{Service Placement Strategies}
\label{sec:service_placement_approaches}

Each algorithm is characterized by its unique strategy for balancing factors such as latency, resource utilization, and \ac{GW} proximity, being the foundation for further enhancements.
This section outlines the core logic of each algorithm, presented through pseudocode, to illustrate their operational mechanisms and provide a basis for comparing their effectiveness in different cloud-to-edge continuum scenarios. 
\textit{Note: Throughout this article, when \ac{BW} is referenced, it is used as the reciprocal ($1/BW$) in the algorithm metrics, providing a more nuanced evaluation of network performance.}

\subsection{Greedy Latency}
\label{subsec:greedy-latency}
    This strategy is presented in Algorithm \ref{alg:greedyLatency}, and it aims to allocate the services in the nodes, sorted by ascending order of the average latency of each node's links, as depicted in Eq. \ref{eq:greedyLatEq}. The performance of this strategy is expected to be one of the best in achieving lower latencies, as the metric used for the sorting of the nodes takes into account the latency of its links. In Algorithm \ref{alg:greedyLatency}, in line \ref{line:greedy_latency_sort_services}, this sorting is what sets apart the two approaches, app-based and service-based. In both strategies, the apps are arranged by decreasing the order of user request rate of each one. However, a distinction emerges in the subsequent sorting of the services. In the app-based approach, the services of the same app remain together, contrary to what happens in the service-based, where the services are ordered in a \ac{RR} process, alternating between all the apps.

    \begin{eqnarray}
\label{eq:greedyLatEq}
	Latency = PD + 1/BW
\end{eqnarray}
\begin{algorithm}[t]
	\caption{Greedy Latency} 
        \label{alg:greedyLatency}
	\begin{algorithmic}[1]

        \State $S \gets$ sort services \label{line:greedy_latency_sort_services}
        \State $N \gets$ sort nodes (ascending order of average latency of each node's links)
        \For{$s$ in $S$}
            \For{$n$ in $N$}
                \If {$ n[FRAM] \geq s[RAM]$}
                    \State place $s$ in $n$
                      \State break
                \EndIf
            \EndFor
        \EndFor

 	\end{algorithmic} 
\end{algorithm}

\subsection{Greedy Free RAM}
\label{subsec:greedy-freeRAM}
    This strategy, presented in Algorithm \ref{alg:GreedyFRAM}, aims to allocate the services in the node with the highest free RAM from the layer closer to the edge available at each iteration.
    Bearing in mind that while this design does not directly make use of link latency-related metrics, the preference for lower layers promotes the physical proximity to the end-user, which may prevent a rise in latency. Additionally, given the fact that it promotes a better resource distribution of services across the network, it is predicted to have an overall good performance, but not the best in decreasing the latency.
    In Algorithm \ref{alg:GreedyFRAM}, in line \ref{line:greedy_FRAM_sort_services}, the procedure is the same applied in algorithm \ref{alg:greedyLatency}, line \ref{line:greedy_latency_sort_services}. Furthermore, in line \ref{line:greedy_FRAM_heapq} the closer the node's layer is to the end-user, the higher its tier is.
\begin{algorithm}[t]
    \caption{Greedy FRAM}
    \label{alg:GreedyFRAM}
    \begin{algorithmic}[1]
        \State $S \gets$ sort services \label{line:greedy_FRAM_sort_services}
        \State nodes\_heap $\gets$ heapq with nodes ordered by ``higher tier'' ``higher FRAM'' \label{line:greedy_FRAM_heapq}
        
        \For{$s$ in $S$}
            \While{$s$ not placed}
                \State $\text{node\_to\_test} = \text{nodes\_heap.pop()}$
                \If {$\text{node\_to\_test[FRAM]} \geq s[\text{RAM}]$}
                    \State place $s$ in $\text{node\_to\_test}$
                    \State $\text{node\_to\_test[FRAM]} -= s[\text{RAM}]$
                    \State $\text{nodes\_heap.push(node\_to\_test)}$
                \Else
                    \State store $\text{node\_to\_test}$ in $\text{nodes\_retrieved}$
                \EndIf
            \EndWhile
            \If{$\text{nodes\_retrieved}$ has nodes}
                \State $\text{nodes\_heap.push(nodes in nodes\_retrieved)}$
            \EndIf
        \EndFor
    \end{algorithmic}
\end{algorithm}

\subsection{Near Gateway}
\label{subsec:near-gw}

In Algorithm \ref{alg:nearGW}, the objective is to achieve an equitable distribution service allocation in proximity to end-users. For this purpose, in line \ref{line:near_gw_sort_services}, the services of the application are sorted based on their message routing sequence. The first service of each application is called the 'zero-order' service, which receives the users' requests. The zero-order services are then allocated to the node with the minimum average distance to the \acp{GW}, which is set as the parent node. The node with the next order services is the one with a lower distance to the parent, and this node becomes the parent to the next order services.
The determination of the distance either to the \acp{GW} or to the parent node is calculated in two distinct ways, which according to the configurations of parameters, set two particular placement strategies: \ac{PD}, depicted in Eq. \ref{eq:nearPDeq} and $\ac{PD}+\ac{BW}$, resorting to the same latency formula portrayed by Eq. \ref{eq:greedyLatEq}. 

According to the prior presented algorithms, this one also differentiates the two approaches: for the service-based approach the message routing sequence is traversed as pictured in lines \ref{line:near_gw_for_order} and \ref{line:near_gw_for_app} of the respective pseudocode; in app-based approach, the message routing sequence is traversed in a reversed progression (i.e., first by apps, and subsequently by order, switching lines \ref{line:near_gw_for_order} and \ref{line:near_gw_for_app} order.

Overall, this algorithm is expected to have one of the best performances in decreasing latency, owing to the primary objective of achieving an equitable distribution of service allocation in proximity to end-users. This is attained by allocating zero-order services closer to the \acp{GW} subsequent services being deployed in the same order as the message workflow and in proximity to the parent node, grouping services closer to the \acp{GW}. While this may leverage the benefits of swarming, reducing overall latency by minimizing both network and geographical distances to the \acp{GW}, it is important to consider the nuances introduced by the modules' internal structures. For instance, scenarios with short initial requests followed by extended message exchanges between services or with end-users may impact latency differently.

\begin{eqnarray}
\label{eq:nearPDeq}
	dist = PD
\end{eqnarray}

\begin{algorithm}[t]
	\caption{Near\_Gateway}
        \label{alg:nearGW}
	\begin{algorithmic}[1]

        \State $M \gets $ Message Routing Sequence dictionary \label{line:near_gw_sort_services}
        \For{$order$ in $M$} \label{line:near_gw_for_order}
            \For{$app$ in $Apps$} \label{line:near_gw_for_app}
                \If {$ order = 0$}
                    \State $n \gets $ node with minimal average $dist$ to GWs
                    \State place $zero\text{-}order\ service$ in $n$
                \Else
                        \For{$s$ in $M[order][app\_index]$}
                            \State $n \gets $ node with minimal $dist$ to parent
                            \State place $s$ in $n$
                        \EndFor
                \EndIf
            \EndFor
        \EndFor

 	\end{algorithmic} 
\end{algorithm}

\subsection{Round-robin IPT}
\label{subsec:RR-IPT}

    Algorithm \ref{alg:RR-IPT} resorts to network partition in communities, utilizing the Louvain graph partition algorithm \cite{10317891} with the \ac{IPT} of each node as metrics, followed by the sorting of communities, and also the nodes within each community, by descendant order of \ac{IPT}. In line \ref{line:RR_IPT_sort_apps}, the apps are sorted by descending average instructions of messages between their services. Consecutively, the sorting of the services is performed, in line \ref{line:RR_IPT_sort_services}, setting apart the two approaches: In app-based, the services of the same app remain together; in service-based, the services are ordered in a \ac{RR} process, alternating between all the apps, as seen in algorithms \ref{alg:greedyLatency} and \ref{alg:GreedyFRAM}. For the placement phase, the services are sequentially allocated in a community, where the nodes are alternated in a \ac{RR} way. If a community does not have resources left for the allocation of a service, that service is allocated to the next community.
    This strategy is presupposed to lower system latency, primarily by deploying apps with a higher average of message instructions in communities with higher \ac{IPT} capacity, aiming to optimize the speed of message processing, thus lowering latency, even though link-associated latency is not taken into account, presumably being one of the algorithms with lower latency.

\begin{algorithm}[t]
	\caption{Round-robin IPT} 
        \label{alg:RR-IPT}
	\begin{algorithmic}[1]
            \State generate louvain communities
            \State sort community nodes by IPT
            \State $Comms \gets$ sort communities by average IPT
            \State $A \gets$ sort apps by average instructions of messages  \label{line:RR_IPT_sort_apps}
            \State $S \gets$ sort services \label{line:RR_IPT_sort_services}
            \State $n \gets 0$
            \For{$s$ in $S$} 
                \For{$c$ in $Comms$}
                    \While{$s$ not allocated}
                        \If {$ n[FRAM] \geq s[RAM]$}
                            \State place $s$ in $n$
                            \State $s \gets$ next service
                        \EndIf 
                        \State $n \gets$ next $c$'s node
                    \EndWhile
                \EndFor
                \If{$s$ not allocated}
                    \State place $s$ in $cloud$
                \EndIf
            \EndFor
	\end{algorithmic} 
\end{algorithm}

\section{Results}
\label{sec:res}
This section presents the findings of a comprehensive study on service placement, mainly within the fog and edge layers of the cloud-to-edge continuum.

The main objective of this investigation is to analyse and compare the performance of app-based and service-based placement approaches in the cloud-to-edge continuum, to appraise their impact on performance, with a focus on minimizing latency in microservices applications.

The proposed methodology, which includes both application-based and service-based placement, is evaluated based on its impact on latency. Furthermore, the distribution of microservices across the network has been analyzed to provide insights into the effectiveness of the service placement approaches. Overall, the study results provide valuable information for optimizing service placement in fog and edge computing environments to improve performance and enhance user experience.

\begin{figure}
    \centering

    \begin{subfigure}{.5\textwidth}
        \centering
        \includegraphics[width=1\linewidth]{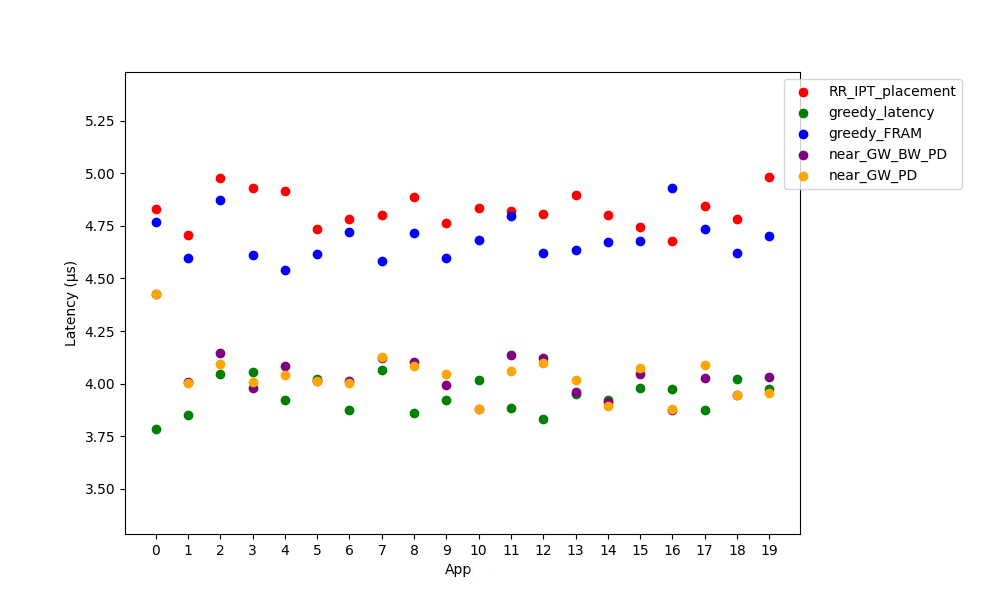}
        \caption{App-Based}
        \label{fig:scatterapp}
    \end{subfigure}
    \begin{subfigure}{.5\textwidth}
        \centering
        \includegraphics[width=1\linewidth]{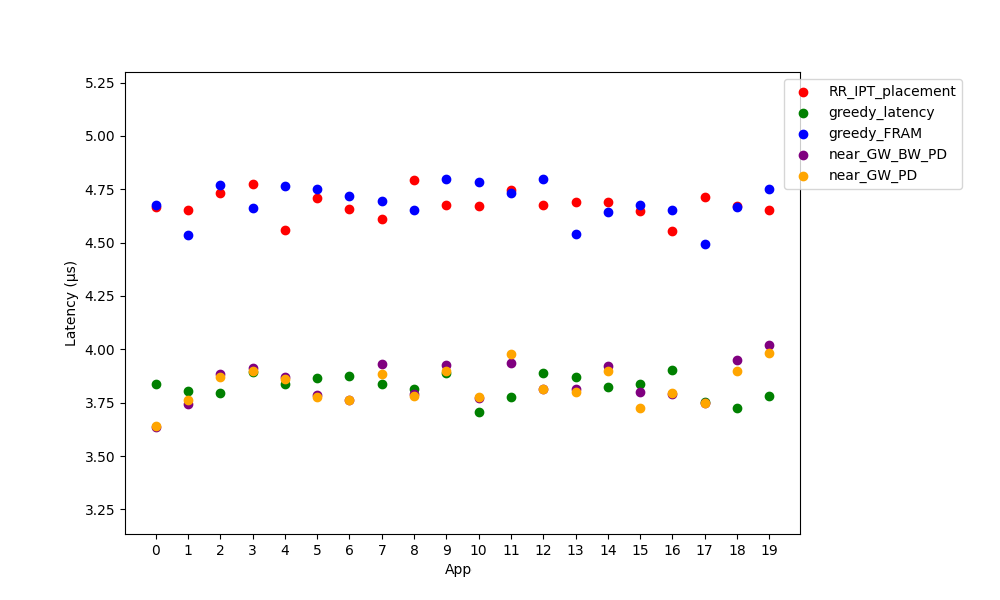}
        \caption{Service-Based}
        \label{fig:scattermod}
    \end{subfigure}%
    \caption{Average App Latency per Algorithm}
    \label{fig:scatter}
\end{figure}

\begin{figure}
    \centering
    \begin{subfigure}{.5\textwidth}
        \centering
        \includegraphics[width=1\linewidth]{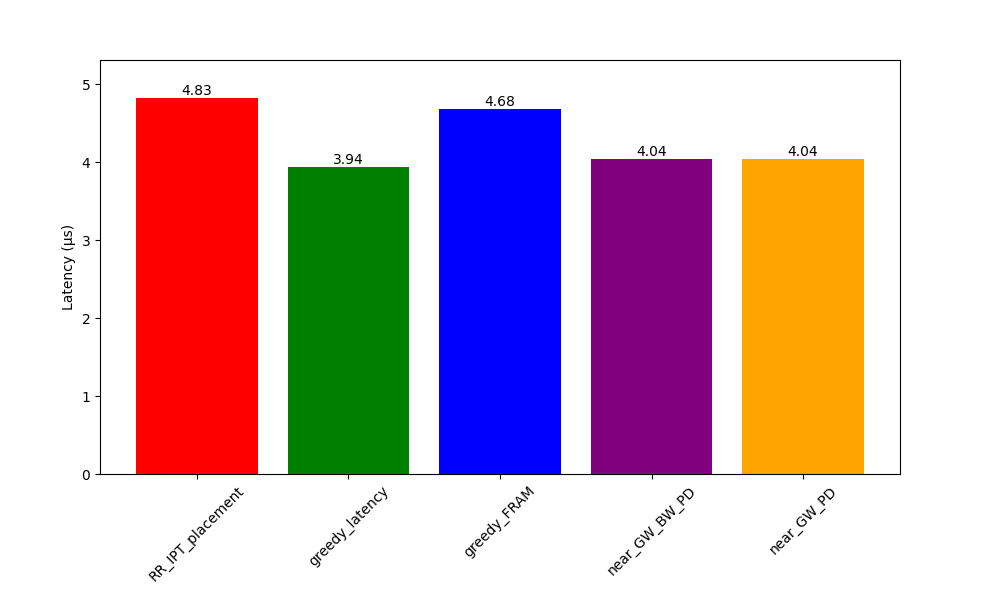}
        \caption{App-Based}
        \label{fig:barapp}
    \end{subfigure}
    \begin{subfigure}{.5\textwidth}
        \centering
        \includegraphics[width=1\linewidth]{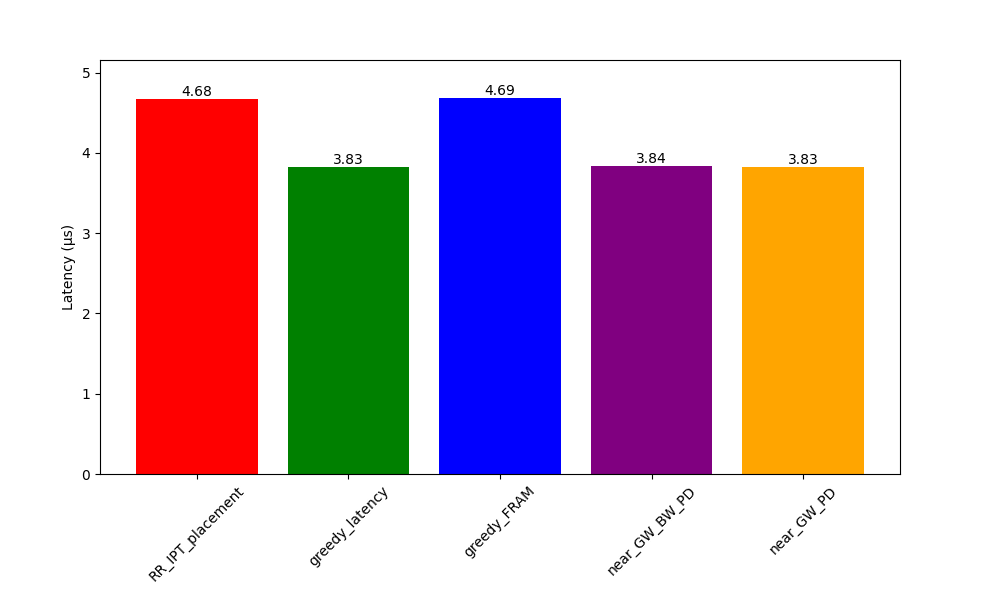}
        \caption{Service-Based}
        \label{fig:barmod}
    \end{subfigure}%
    \caption{Average Latency per Algorithm}
    \label{fig:barplot}
\end{figure}

Results for the latency values obtained through the simulation of different algorithms are presented in Figures \ref{fig:scatter} and \ref{fig:barplot}. Both figures display the results of app-based (\ref{fig:scatterapp} \& \ref{fig:barapp}) and service-based (\ref{fig:scattermod} \& \ref{fig:barmod}) approaches.
Figure \ref{fig:scatter} shows the mean latency per application across all algorithms; similarly, Figure \ref{fig:barplot} gives a more simplified view of the average latency for the algorithms, allowing for easier comparative analysis. After analyzing both Figures (\ref{fig:scatter} and \ref{fig:barplot}), it can be concluded that Algorithms \ref{alg:greedyLatency} (Greedy Latency) and \ref{alg:nearGW} (Near GW), with both metrics, perform better than the others, in terms of average latency in both scenarios. It was also observed that there were differences in performance between the two approaches (Figure \ref{fig:barapp} \& Figure \ref{fig:barmod}). The service-based approach (Figure \ref{fig:barmod}) had a lower average latency in most cases, between $0.11\mu s$ and $0.20\mu s$. The exception is Algorithm \ref{alg:GreedyFRAM} (Greedy Free RAM), which did not show any significant differences between the two scenarios.

Along with the latency results, the load balance of the network is also meaningful. These results are presented in Table \ref{tab:piechart}, which features the percentage of nodes used by each algorithm, in both approaches (a node is considered occupied if it has at least one microservice allocated). From this analysis, it is noticeable a better load distribution when testing the simulation with Algorithms \ref{alg:GreedyFRAM} (Greedy Free RAM) and \ref{alg:RR-IPT} (Round-robin IPT). Moreover, the algorithms have a similar distribution of load between app-based and service-based approaches.

\begin{table}[t]
\captionof{table}{Percentage of Used Nodes per Algorithm}
\centering
\begin{tabular}{|c|c|c||c|c|}
\hline
\multirow{2}{*}{Algorithm} & \multicolumn{2}{c||}{App-Based} & \multicolumn{2}{c|}{Service-Based} \\
\cline{2-5}
& \% Used & \% Unused & \% Used & \% Unused \\
\hline
RR\_IPT\_placement & 17.0 & 83.0 & 17.4 & 82.6 \\
\hline
Greedy\_Latency& 12.8 & 87.2 & 12.8 & 87.2 \\
\hline
Greedy\_FRAM& 22.7 & 77.3 & 22.6 & 77.4 \\
\hline
Near\_GW\_BW\_PR & 16.7 & 83.3 & 16.3 & 83.7 \\
\hline
Near\_GW\_PR & 16.6 & 83.4 & 16.4 & 83.6 \\
\hline

\end{tabular}
\label{tab:piechart}
\end{table}

\section{Discussion}
\label{sec:disc}

In the case of Algorithm \ref{alg:greedyLatency} (Greedy Latency), as expected, the prioritization of nodes whose links had lower average latency clearly benefits the application in achieving the lowest simulation latency of this study, pictured in Figures \ref{fig:scatter} and \ref{fig:barplot}.

Meanwhile, with Algorithm \ref{alg:nearGW} (Near Gateway), the strategy of deploying services as a swarm near the end-users exhibits the 2\textsuperscript{nd} best latency performance when using $\ac{BW}+\ac{PD}$ or just \ac{PD} as metrics, achieving the objective of minimizing latency. The similarity in both performances may be due to the magnitude of contrast between \ac{BW} and \ac{PD}. Given the fact that \ac{PD} outweighs \ac{BW} when including the latter to calculate the distance, the difference is negligible, not conferring a significative increase in performance. 
Comparatively to Algorithm \ref{alg:greedyLatency} (Greedy Latency), the current strategy has a better load balance, while still having close latency values. 

The analysis of Algorithm \ref{alg:GreedyFRAM} (Greedy Free RAM) endorses the particular forecasted outcome of this strategy. Bearing its characteristic of spreading the load across the topology while still giving priority to lower layers' nodes, but not taking link attributes into account, it leads to the highest node occupancy, pictured in Table \ref{tab:piechart}, and to be one of the worst performant algorithms. The fact that this strategy relies solely on node resources, neglecting latency corresponding to link attributes, is the potential reason for no disparity between the two placement approaches.

Lastly, Algorithm \ref{alg:RR-IPT} (Round-robin IPT), based on communities partition of the network, does not accurately reflect the prediction that by allocating apps with a higher average of message instructions in higher \ac{IPT} capacity communities, the system latency would be one of the lowest. In fact, it has one of the worst latency performances, in pair with Algorithm \ref{alg:GreedyFRAM} (Greedy Free RAM) in the service-based approach. This disparity from what was expected, can possibly be due to the greater spread of services across the network, mainly by performing a \ac{RR} around the whole community, placing services in nodes farther from the end-user, thus increasing latency. On the other hand, this characteristic enhances service distribution in the continuum.

Table \ref{tab:piechart} displays the distribution of application services across various network nodes. Based on the analysis of the results, it can be inferred that a greater dispersion of nodes, i.e., having more nodes in the network with at least one service allocated, is associated with poorer performance in terms of latency, although it may promote better resilience. This can be explained by the fact that if services are spread out, communication may have to travel a greater distance to ensure communication within the app services, which results in slower response time. On the other hand, a lesser distribution of services ensures that the services are closer to each other, resulting in faster communication.
Furthermore, the minimal differences obtained between the percentage of used nodes in both scenarios (App-Based and Service-Based) can be explained by the fact that the placement algorithms only change the order in which services are allocated. As a result, the decisions related to the selection of a network node remain unaffected.

Overall, aligned with the main objective of this investigation, as highlighted in Table \ref{tab:better_approach}, placement strategies with a service-based approach have achieved better latency performance in 4 of the 5 algorithms used and a draw in the remaining one. Thus, evidencing the advantage of a more uniform distribution of each app's services over the network as its load intensifies, benefiting a more homogeneous latency performance across all applications.

\begin{table}[t]
\captionof{table}{Lower Latency Approach per Algorithm}
\centering
\begin{tabular}{|l l|c|}
\hline
\multicolumn{2}{|c|}{Algorithm} & Lower Latency Approach \\
\hline
(\ref{alg:greedyLatency})&Greedy Latency & service-based \\
\hline
(\ref{alg:GreedyFRAM})&Greedy FRAM & draw \\
\hline
(\ref{alg:nearGW})&Near GW BW+PD & service-based \\
\hline
(\ref{alg:nearGW})&Near GW PD & service-based \\
\hline
(\ref{alg:RR-IPT}) & RR IPT & service-based \\
\hline

\end{tabular}
\label{tab:better_approach}
\end{table}

\section{Conclusion}
\label{sec:conc}

As hypothesized, either of the strategies employed through the algorithms depicted in Section \ref{sec:service_placement_approaches} performed better or equally when resorting to the service-based placement approach, compared to the app-based approach. The distinctive principles behind each strategy: link latency, resources load balance, \ac{GW} proximity or community partition and processing capacity; while getting consistently better results with a service-based approach, support the thesis elaborated that a service-based design is better qualified to achieve lower response time in cloud-to-edge scenarios.

Furthermore, the comprehensive analysis conducted using \ac{YAFS} not only confirms the superiority of the service-based approach but also highlights the practical implications of each placement algorithm. The results obtained contribute to the ongoing discourse on the optimization of microservices placement in the cloud-to-edge continuum, offering a deeper understanding of the challenges and opportunities in real-time data processing scenarios.

Future research in this domain could explore further refinements of placement algorithms, namely evolutionary ones, and investigate additional criteria and goals for optimization. Additional implications on privacy when relying on edge and fog nodes could also be explored, as well as network resilience. Finally, extending the evaluation to diverse network topologies and workload scenarios would enhance the generalizability of the current findings. Overall, this work lays the groundwork for continued exploration and innovation in the dynamic field of fog and edge computing.

\section*{Acknowledgement}
\label{sec:ackn}
\footnotesize
This work is funded by the FCT - Foundation for Science and Technology, I.P./MCTES through national funds (PIDDAC), within the scope of CISUC R\&D Unit - UIDB/00326/2020 or project code UIDP/00326/2020; and by the project POWER (grant number POCI-01-0247-FEDER-070365), co-financed by the European Regional Development Fund (FEDER), through Portugal 2020 (PT2020), and by the Competitiveness and Internationalisation Operational Programme (COMPETE 2020).

\bibliography{ref}

\end{document}